\documentclass[useAMS,usenatbib,usegraphicx]{mn2e}
\usepackage{epsfig}

\title[Observable constraint on the frequency-mass relation of compact stars]{Observable constraint on the frequency-mass relation of compact stars}

\author[X. P. Zheng, N. N. Pan, L. Zhang]{X. P. Zheng\thanks{Email: zhxp@phy.ccnu.edu.cn}, N. N.
Pan\thanks{Email:Pannana@phy.ccnu.edu.cn}, L. Zhang\\
Institute of Astrophysics, Huazhong Normal University, Wuhan
430079, P. R. China}

\begin{document}
\maketitle

\begin{abstract}
We study the relation between mass and
limit rotation for various kinds of compact-star models. Under the
considerations of Keplian
 and r-mode instability limits, we calculate tens of the limit frequency-mass relation curves.
We present constraint of the observable rotation frequency on the
frequency-mass relations. Along with the observable constraint on
mass-radius relation, we may actually probe into the interior of
compact stars. Our results show that the possible submillisecond
pulsars would be serious to distinguish the compact stars
contained quark matter from normal neutron stars. The recent
discovery of burst oscillation at 1122Hz in the X-ray transient
XTE J1739-285 provides us a challenge. As an example, the
constraints at 1122Hz level allow the hybrid-star radii in the
range $9{\rm km}\leq R\leq 12{\rm km}$ and the  masses in the
range $1.2M_\odot\leq M\leq 2.0M_\odot$.
\end{abstract}

\section{Introduction}
The interior of compact stars contains nuclear matter at
very high densities that are a few times and even up to more than
ten times the density of ordinary atomic nuclei. It provide a
high-pressure environment in which numerous subatomic particles
compete with each other. In early years, compact stars were
regarded as the neutron stars which are constituted of neutrons
and protons. However nuclear physical theory and experiments favor
the presence of hyperons and the Bose condensates of pions and
koans in a few times the equilibrium density of nuclear matter.
The quantum chromodynamics theory also predicts the occurence of
deconfinement phase transition in such high density. A large
variety of particles other than just neutrons and protons could
extensively exist inside various compact stars.

It has so far proved very difficult to find any effective method
that could unambiguously distinguish them. Commonly ones believe
that the maximum masses and typical radii of compact stars are the
two most important properties which reflect rather difference
among the equations of state(EOSs) of dense matter. The pioneers
hereby presented the observable constraints on mass-radius
relations. The maximum mass of compact stars has a limit of
$3M_\odot$, which is a consequence of general relativity(Rhoades
Jr.\& Ruffini 1974). The maximum mass is also controlled by the
stiffness of dense matter EOS at densities in excess of a few
times saturation density. The inclusion of non-nucleonic degrees
of freedom at supra-nuclear densities generally implies a
softening of the EOS. Some recent mass measurements, from timing
of pulsars in binaries with white dwarf companions, observations
of quasi-periodic oscillations and x-ray emission of accreting
neutron stars, seem to suggest a maximum mass in the vicinity of
$2M_\odot$(Nice et al 2005) but the precisely known
 mass is only $1.44M\odot$(Taylor\& Weisberg 1989), or about $1.68M_\odot$ at 95 \%
 confidence(Ransom et al, 2005). As for compact star radii, the measurements are far
 less precise than mass measurements. These masses and radii are
 unable to provide a good constraints on the compositions of dense
 matter and the structural properties of compact stars because
 most EOSs are allowed by the measurements.

Aside from direct mass and radius determinations, rapid rotation
 is another crucial property of compact stars. Measurements of
 pulsar spin frequency are extremely precise. Nearly two thousands
 of pulsar frequencies are well measured, tens of which have
 frequencies over 100Hz. The first millisecond pulsar spinning at
 641Hz was discovered in 1982(Backer  et al). And then a new one was reported
 during the next decade or so almost every year. The fastest presently known
 pulsar has 716Hz spin frequency which was reported in
 2006(Jason et al). These are perhaps utilizable for constraining
 compact stars. The most well known attempt is the out of critical
 mass-radius relation by the measured large frequency assumed  as
 Keplerian limit(Glendenning 1997; Lattimer\& Prakash 2007). This way, displayed below, is not reliable to
 exclude the compositions of dense matter. The Keplerian frequency
 cannot reach in many situations. It is because other
 instabilities inside compact stars are even more efficient in
 determining the largest rotations. Some researches found the
 r-mode instability would play a key role for many cases.
 Immediately it has been realized  that compact star spin
 evolution is also sensitive to the internal compositions which
 determine or not the dissipations of unstable modes are large or
 small. Ones were impelled to probe the interior of compact stars
 through the measured rotations restricting the instability
 windows. The efforts from this aspect is sometimes fruitless yet.
 In the circumstances, we think of the requirement of a new scheme for the
 stringent constraints on compact stars through pulsar's
 frequencies. We suggest (1) the frequency-mass relations for
 various EOSs are constrained by the measured pulsar rotation frequencies, different from the conventional
 considerations of constraints on mass-radius relations. (2) Both
 Kepler and r-mode limit rotations are taken into account
 together.

 We will below study various compact-star models, normal neutron
 star(NS), strange star(SS), hyperon star(HpS) and hybrid star(HbS, the inclusion of Maxwell(MC) and Gibbs(GC)
 constructions). We apply RMF(Glendenning 1997), APR(Akmal et al 1998) and
 BBG( Nicotra et al 2006) for NSs, MIT and eMIT(Schertle et al 1997) for SSs,
 RMF(Lackey et al 2006), BBG(Baldo et al 2000) for HpSs, RMF+MIT(Pan\& Zheng 2007),
 RMF+eMIT(Schertle et al 2000; Zheng et al 2007) and BBG+MIT(Fahri\& Jaffe 1984; Nicotra et al 2006)for
 HbSs. This paper is arranged follows as. We discuss the viscous
 dissipations and estimate that of the interface between pure
 quark matter and pure hadronic matter in section II. We
 investigate the  constraints of the rotations on frequency-mass
 relations for  compact stars in section III. We present our
 conclusions and discussions in section IV.

 \section{Viscous dissipation in compact-star matter}

Besides shear viscosity, bulk viscous dissipation is one of the important factors that determines
 the limit rotation of a
 compact star.  Up to now, the viscosities of homogeneous neutral phases and
  the mixed hadron-quark phases by bulk Gibbs calculation have been extensively
 studied. The considerations of the dissipation in various dense matter have been listed in Table 1. In addition,
the surface rubbing of compact stars with solid crust has been
also calculated(Bildsten  and Ushomirsky 2000; Andersson  et al,
2000).  However, the ¡°bulk Gibbs¡± is too simple to study the
mixed phase, since we must consider matter with non-uniform
structures instead of two bulk uniform matters; the mixed phase
should have various geometrical structures where both the number
and charge densities are no more uniform(Endo et al 2006). Then we
have to consider finite-size effects like the surface and the
Coulomb interaction energies. Unfortunately we can't determine the
parameters for surface tension and charge screening. Endo et
al(2006) showed that we shall see EOS results in being similar to
that given by MC under extreme condition. So we regard GC and MC
as upper and lower limits of mixed phase. For a MC HbS, the
dissipation in the transition interface between quark matter core
and hadron matter envelope needs to be estimated except two
homogeneous neutral phases. We here have to reckon with the
dissipation in the interface between pure quark matter and pure
hadronic matter.

\begin{table}
\caption{Bulk viscosities in various dense matter. Compositions
refers to the interacting components, n - neutrons, p - protons, e
- electrons, H - hyperons and Q - quarks. } \centering
\begin{tabular}{c|c|c}
\hline\hline  Compositions   &   References
&  Adaptive models \\
\hline
 npe   & Sawyer (1989)   &  NS, HpS, HbS \\
 npeH   & Jones (2001)   &  HpS \\
  npeQ    & Drago et al (2005)  & HbS \\
   npeHQ    & Pan \& Zheng  (2006)  & HbS \\
   Q(u,d,s)    &  Madsen (1992) & SS, HbS \\
\hline \hline
\end{tabular}
\label{tab:data}
\end{table}

 Let us assume that the HbS undergoes a radial pulsation
$\delta n=\Delta n \sin {2\pi\over\tau}t$. During the course, the
hadrons penetrate the interface to be transformed into quarks when
$\delta n>0$ in $(0, {\tau\over 2})$, while quarks confine into
hadrons when $\delta n<0$ in $({\tau\over 2},\tau)$. Using the
standard definition of the bulk viscosity and the dissipation of
the energy in the hydrodynamic motion due to the irreversibility
of  periodic compression-decompression process, we can write the
energy dissipation rate per unit volume averaged over the
pulsation period $\tau$  as, $
\left\langle\dot{\varepsilon}\right\rangle=-\zeta\left\langle({\rm
div}{\bf v})^2 \right\rangle$ and $
\left\langle\dot{\varepsilon}\right\rangle=-{n\over\tau}\int^\tau_0
dt {d\over dt}\left ({1\over n}\right )$.

Supposing the baryon densities for
the quark matter phase and hadron matter
 phase on
both sides of the interface are $n_{\rm Q}$ and $n_{\rm H}$, we
can get
\begin{equation}
\zeta\left\langle({\rm div}{\bf v})^2 \right\rangle={(n_{\rm
Q}-n_{\rm H})^2\over n_{\rm Q}n_{\rm H}}{P_{\rm tran}\over\tau}.
\end{equation}
Thus the damping on pulsations due to confined-deconfined
processes
 in the transition boundary layer is obtained as
\begin{equation}
\left({dE\over dt}\right )_{\rm tran}=- {(n_{\rm Q}-n_{\rm
H})^2\over n_{\rm Q}n_{\rm H}}{P_{\rm tran}l_{\rm tran}4\pi
R^2_{\rm tran}\over\tau},
\end{equation}
where $P_{\rm tran},l_{\rm tran}$ and $R_{\rm tran}$ represent the
pressure in the interface, the layer thickness and the radius of
quark matter core inside the MC HbS respectively. The ¡°height¡±
of the surface waves induced by the r-modes has been given by
${\rm\delta}r_s\approx 0.07\alpha\left
({\Omega\over\Omega_K}\right )^2$ in dimensionless(Kokkotas \&
Stergioulas 1999). We then estimate the layer thickness inside the
HbS by $l_{\rm tran}/R\equiv{\rm\delta^{(2)}}r\leq
({\rm\delta}r_s)^2\sim 10^{-4}\alpha^2\left
({\Omega\over\Omega_K}\right )^4$, where $R$ denotes the radius of
the star. We immediately estimate $l_{\rm tran}\leq 10$cm for
$R=10$km.

\section{Constraints on frequency-mass relations}
For a uniform rigid sphere with mass M and radius R, the
mass-shedding limit is(Lattimer \& Prakash 2004)
\begin{equation}
{1\over\nu_{\rm K}}=0.96{(R/10{\rm km})^{3/2}\over
(M/M_\odot)^{1/2}},
\end{equation}
and the critical rotation frequency for a given stellar model can
also be derived from following equations
\begin{equation}
{1\over\tau_{\rm gr}}+{1\over\tau_{\rm v}}=0,\ \ \ \ \nu_{\rm
R}=\min{[\nu(T)]},
\end{equation}
where $\tau_{\rm gr}<0$ is the characteristic time scale for
energy loss due to the gravitational waves emission, $\tau_{\rm
v}$ denotes the damping timescales due to the shear, bulk
viscosities and other rubbings. Surface rubbing is decisive for
NSs,HpSs and HbSs, whereas ${1\over\tau_{\rm sr}}$ can be
neglected even for quark stars with maximal crust. The shear
viscosity timescale for strange quark matter is calculated
by(Heiselberg and Pethick 1993). We here treat the bulk viscosity
timescale with Lindblom et al method(1999). In order to constrain
neutron-star matter EOS, as well-known, the traditional treatments
of (3) and (4) are done in past works. From (3), a series of $M,
R$ upper limits are often obtained if $\nu_K$ is surely given,
such as 716Hz of the fastest rotation pulsar. The inferred $M-R$
curve can be used to confine the mass-radius relations of various
compact stars and hence the EOSs. On the other hand, the equation
(4) can give r-mode instability windows in temperature-frequency
plane for the given compact stars. In comparison to the measured
frequencies, the windows may pinpoint to the possible compact-star
models.

However we below will see the traditional treatments are not
always dependable. We  will give up the previous idea.  As we
known, a compact star sequence can be constructed for a given EOS.
Let the compact star sequence as the input of equations (3) and
(4), we immediately obtain the corresponding frequency-mass
relations, $\nu_{\rm K}-M(R)$ and $\nu_{\rm R}-M(R)$.
Consequently we can find a genuine upper frequency from taking
extreme value $\nu_{\rm C}=\min(\nu_{\rm K}, \nu_{\rm R})$  for a
compact star with $M$ (and associated $R$). We impose the measured
frequencies on the frequency-mass relations and hence tightly
constrain the $\nu_c-M$ relations.

Fig 1  shows the $\nu-M$ relations on the basis of eqs (3) and (4)
for typical NS, HpS, HbS(including MC and GC constructions) and SS
sequences. They represent the stars which have the maximum
rotation frequencies in respective models. Once the rapid
rotations of pulsars are measured we can constrain the theoretical
$\nu-M$ relations with the rotational frequencies. The observed
millisecond pulsars are aimed at this issue. Figure 1 means that
the observations only favor the models satisfying the condition
$\nu_{\rm obs}\leq\nu\leq\nu_c(M)$. The most models for HpSs and
HbSs are compatible with the fastest rotating pulsar found yet,
J1748-2446ad. The situation will be very different if higher
frequency is regarded as the confined condition. The pulsar over
1000Hz frequency is perhaps decisive. XTE J1739-285 in which
Kaaret reported the discovery of a burst oscillation at
1122Hz(Kaaret et al 2007) may be considered a candidate of such
pulsar. Above 1122Hz, figure 1 shows that most models are ruled
out. Some of only HbSs are possible. In comparison, we also have
plotted the corresponding mass-radius relations in Fig. 2. Check
Fig 2 against Fig 1, we find the constraints of observable masses
on $M-R$ relations, even a supplement of traditional treatment of
Keplerian limit(XTE J1739-285 in Fig2) to that, is rather wide or
not always dependable. Fig 1 also hints that the traditional
treatment of r-mode instability window plus Keplerian
limit(defined as a fixed value such 1122Hz in Fig1) supports the
models with limit frequencies above the fixed value but no
available(for example, the dashed curves above 1122Hz but out of
the dark green region).
\begin{figure}
\includegraphics[width=0.52\textwidth]{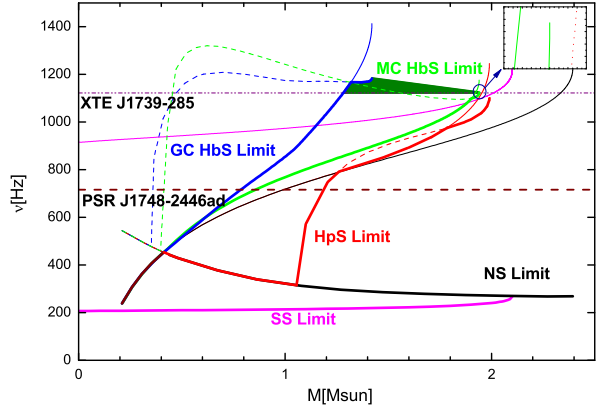}
\caption{The constraint of the observable frequencies on
frequency-mass relations for various models.
  The corresponding thick lines are the
minimum frequencies for various compact star sequences that are
taken in our calculations. The inserted figure here is a
amplificatory image of the local. We find that most models are
allowed by 716Hz rotation of J1748-2446ad but almost all the
models couldn't reach the limit by the 1122Hz of XTE J1739-285
expect the HbS model, which is denoted by the dark green region.
The possible range of limit frequencies
 is so narrow that the possibility for the existence of compact stars with submillisecond period is
  small. } \label{fig:mf}
\end{figure}

\begin{figure}
\includegraphics[width=0.52\textwidth]{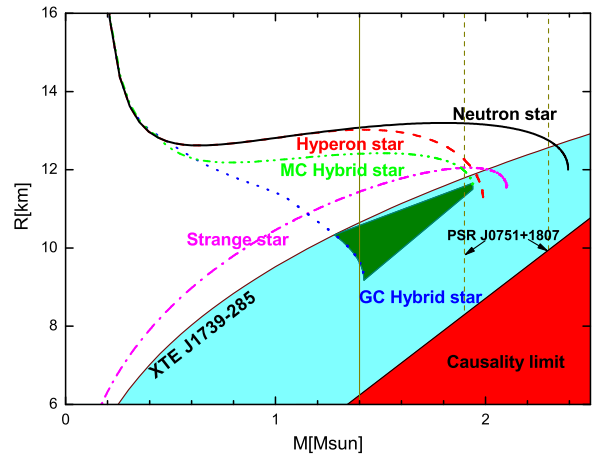}
\caption{The corresponding  mass-radius relations of compact stars
imposed by observations. The dark green contour represents our
tight constraint of the masses and radii from Fig 1, which is
evidently allowed by mass and causality limits. Although some of
NSs, HpSs and SSs are also permitted by mass  and simply
traditional Keplerian limits denoted by the cyan district, which
has been used in past works, they are excluded by the genuine
limit frequency at 1122Hz.} \label{fig:mr}
\end{figure}
\section{Conclusions and discussions}
The EOS and composition of dense matter in the core of compact
stars have attracted much attentions owing to their importance for
our understandings of compact stars. The researchers made great
efforts to distinguish various dense matter suggested by nuclear
physical theory. The most familiar treatment is the constraint of
observable masses on the mass-radius relation of compact stars. We
think it not to be always dependable. We present the constraint of
observable rotation on the frequency-mass relation of compact
stars and hence EOS of the dense matter. Under our considerations,
both Keplerian and r-mode instability limits are involved.The
constraint has been displayed to advantage, more stringent than
that of masses. Especially, the constraint may bring us the
decisive judgement if a submillisecond pulsar would be really
confirmed in future. We have calculated tens of results  but only
the typical cases in which hadronic matter is in description of
RMF are depicted in the figures. There are slight but nonessential
changes when the other EOSs are applied. In addition, although
magnetic fields can be important to suppress r-mode instabilities
due to which the star
 may rapidly slow down other than the
 gravitational wave emission,  their effect is probably negligible for
frequencies exceeding 0.35
 times the mass-shedding limit under the conditions that if the
 field B is not larger than $10^{16} (\Omega/\Omega_{K})$
 Gauss, here $\Omega$ and $\Omega_{K}$ represent the angular velocities
 of the star and the mass-shedding limit respectively(Rezzolla  et al 2000).
 This couldn't do any
 influence on our conclusion about the star rotating above 1000 Hz at all,
 above which the effect of gravitational wave emission is important.
  In all, we argue that
quark matter has perhaps been indicated by rotations in possible
submillisecond pulsars.

This research is supported by the National Natural Science
Foundation of China under Grant No. 10773004. Pan is also
partially supported by the National Natural Science Foundation of
China under Grant No. 10603002.


\begin{thebibliography}{}

\bibitem[1998]{ak98} Akmal A. et al,
Phys. Rev. C,  58, 1804.
\bibitem[2000]{and00} Andersson N.  et al., 2000,
 ApJ, 534, L75.

\bibitem[1982]{Bac82}  Backer D.C. et al.,
1982, Nature,  300, 615.
\bibitem[2000]{Bal00} Baldo M.et al, 2000,
Phys. Rev. C., 61, 055801
\bibitem[2000]{bil00}  Bildsten L. and Ushomirsky  G., 2000,
ApJ, 529, L33.
\bibitem[2005]{dra05}Drago A. et al, 2005, Phys.
Rev. D, 71, 103004.
\bibitem[2006]{end06} Endo  T.  et al.,    2006, {Proceeding of 29th
Johns Hopkins Workshop: "Strong Matter in the Heavens", in
Proceedings of Science, PoS(JHW)019}.
\bibitem[1984]{Fah84} Fahri E.  and Jaffe R. L. 1984,
Phys. Rev. D., 30, 2379
\bibitem[1997]{Gle97} Glendenning N. K. 1997, Compact
stars, Springer-Verlag.
\bibitem[1993]{hei93}  Heiselberg H. and  Pethick C. J., 1993,
Phys. Rev. D, 48, 2916.
\bibitem[2006]{jas06} Jason W.T.   et al.,
 2006, Science,  311, 1901.
\bibitem[2001]{jon01}Jones P. B., 2001, Phys. Rev. D, {\bf 64}, 084003.
\bibitem[2007]{kaa07} Kaaret P.  et al.,
 ApJ, 657, L97.
 \bibitem[1999]{kok99} Kokkotas K.D. and
Stergioulas N., 1999, A$\&$A,  341,
110.
\bibitem[2006]{lac06}Lackey  B.D.  et al., 2006, Phys. Rev D,
73, 024021.
 \bibitem[2004]{lat04} Lattimer J.M. and Prakash M., 2004,
 Science,  304, 536.
\bibitem[2007]{lat07} Lattimer J.M. and  Prakash M.,
 2007, Phys. Rep.,  442, 109.
\bibitem[1999]{lin99} Lindblom L. et al, 1999, Phys. Rev. D, 60, 064006.
\bibitem[1992]{mad92} Madsen  J., 1992, Phys. Rev. D,  46, 3290.
\bibitem[2005] {nic05}Nice D. J. et al, 2005, ApJ, 634,1242.
 \bibitem[ 2006]{Nic06} Nicotra O. E. et al. 2006,
Astron. Astrophys., 451, 213.
\bibitem[2006]{pan06} Pan N.N. and Zheng X. P.,
2006, MNRAS, 371, 1359.
\bibitem[2005] {ran05}Ranson S.M. et al,
2005, Science, 307,892.
\bibitem[2000]{Rez00} Rezzolla L.  et al., 2000,
 ApJ, 531, L141.
\bibitem[1974] {roa74}Roades Jr. C. E. and
Ruffini R, 1974, Phys. Rev. Lett. 32, 324.
\bibitem[1989]{saw89} Sawyer R. F. 1989,
Phys.Lett.B., 233, 412.
\bibitem[1997]{Sch97} Schertler K.,  Greiner C.  and Thoma M. H. 1997,
Nucl. Phys. A., 616, 659.
\bibitem[1989] {tay89}Taylor J. H. and
Weisberg J. M., 1989 ApJ, 345,434.

































\end{thebibliography}
\end{document}